\documentclass[aps,amsmath,amssymb,notitlepage,onecolumn,nofootinbib,floatfix,superscriptaddress]{revtex4-1}

\pdfoutput=1

\usepackage[T1]{fontenc}  

\usepackage[version=3]{mhchem} 
\usepackage{graphicx} 

\usepackage{amsmath} 
\usepackage{amssymb} 

\usepackage{hyperref} 
\usepackage[all]{hypcap}    
\usepackage{pgf} 
\usepackage{array}
\usepackage[top=1in, bottom=1in, left=1in, right=1in]{geometry}
\usepackage{multirow} 
\usepackage{siunitx}
\usepackage{algorithm,algorithmic}
\usepackage{xcolor}

\usepackage{paralist}

\usepackage[caption=false]{subfig}

\usepackage{textcomp}  

\usepackage{fancyhdr}

\newcommand{\us}{\textmu{}s}

\newcommand{\dwave}{\mbox{D-Wave}}  

\newlength\figureheight 
\newlength\figurewidth

\DeclareGraphicsExtensions{.pdf,.png,.jpg}
\graphicspath{{figures/}}

\newcommand{\graphicinput}[2][]{\includegraphics[#1]{#2}}

\newcommand{\hide}[1]{}

\renewcommand{\epsilon}{\varepsilon}

\definecolor{titlecolor}{RGB}{0,51,102}
\hypersetup{
	colorlinks = true,
	allcolors = {titlecolor}
}

\begin{document}

\title{Quantum-Assisted Genetic Algorithm}
\author{James King}
\affiliation{\dwave{} Systems}
\author{Masoud Mohseni}
\affiliation{Google, Inc.}
\author{William Bernoudy}
\affiliation{\dwave{} Systems}
\author{Alexandre Fr\'{e}chette}
\affiliation{\dwave{} Systems}
\author{Hossein Sadeghi}
\affiliation{\dwave{} Systems}
\author{Sergei V. Isakov}
\affiliation{Google, Inc.}
\author{Hartmut Neven}
\affiliation{Google, Inc.}
\author{Mohammad H.~Amin}
\affiliation{\dwave{} Systems}

\date{\today}

\begin{abstract}
Genetic algorithms, which mimic evolutionary processes to solve optimization problems, can be enhanced by using powerful semi-local search algorithms as mutation operators. Here, we introduce reverse quantum annealing, a class of quantum evolutions that can be used for performing families of quasi-local or quasi-nonlocal search starting from a classical state, as novel sources of mutations.  Reverse annealing enables the development of genetic algorithms that use quantum fluctuation for mutations and classical mechanisms for the crossovers---we refer to these as Quantum-Assisted Genetic Algorithms (QAGAs). We describe a QAGA and present experimental results using a D-Wave 2000Q quantum annealing processor. On a set of spin-glass inputs, standard (forward) quantum annealing finds good solutions very quickly but struggles to find global optima. In contrast, our QAGA proves effective at finding global optima for these inputs. This successful interplay of non-local classical and quantum fluctuations could provide a promising step toward practical applications of Noisy Intermediate-Scale Quantum (NISQ) devices for heuristic discrete optimization.
\end{abstract}

\maketitle

\thispagestyle{fancy}

\makeatletter
\def\l@subsubsection#1#2{}  
\makeatother


\parskip=5pt
\parindent=0pt

Successful classical heuristic search algorithms often combine a global search for good regions of the solution space with local refinement of good solutions that have been found---the complementary tasks of \textit{exploration} and \textit{exploitation}. Global search and local refinement are integral to the quantum annealing algorithm \cite{Kadowaki1998}, a metaheuristic that is similar to simulated annealing but uses quantum effects to accelerate computation.  In quantum annealing, the balance between global search and local refinement is controlled by the strength of the transverse field, which is roughly analogous to the temperature parameter in simulated annealing \cite{Kirkpatrick1983}.  In the canonical version of quantum annealing (i.e., forward quantum annealing, or simply forward annealing), early in the anneal when the transverse field is strong, tunneling is frequent and the solution space is searched globally.  Late in the anneal when the transverse field is weak, the algorithm's search narrows in on local regions of the solution space.  A similar transition occurs in simulated classical annealing, wherein a high-temperature random global search early in the annealing algorithm gives way to low-temperature greedy local refinement later on.

While forward quantum annealing takes advantage of global search and local refinement, it does so within a black box---nothing between the initial superposition and the final answer can be seen or controlled by the user due to the no-cloning theorem of quantum mechanics \cite{wootters1982single}.  This means that forward quantum annealing cannot be used to perform local refinement of good candidate solutions found elsewhere, i.e., from a previous anneal or through classical methods.

\begin{figure}[ht]
	\includegraphics[width=4in]{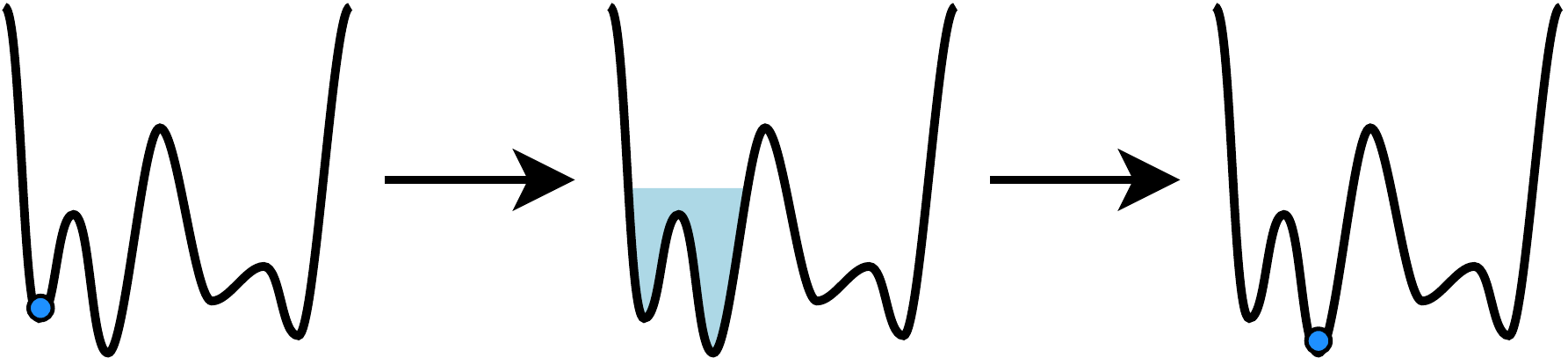}
	\vspace{-3mm}
	\caption{Sketch of the reverse annealing process.  The anneal starts at a specified classical state (left), then performs a quasi-nonlocal quantum annealing search enforced by an increased transverse field (middle), then settles in a new classical state as the transverse field is removed (right).\label{fig:reverse_annealing}}\vspace{-3mm}
\end{figure}

To address this limitation, the D-Wave 2000Q system provided a \textit{reverse annealing} feature, inspired by proposal \cite{PatentGoogle}, enabling the use of quantum critical phenomena for local refinement of classical solutions; making it possible to invoke quantum annealing as a component in more sophisticated hybrid algorithms \cite{DWaveReverseAnnealing}. From a specified classical starting state, a quasi-local or quasi-nonlocal search is stimulated by increasing the transverse field, effectively annealing backward from a classical state to a mid-anneal quantum superposition; after allowing the search to run for a specified amount of time at this mid-anneal point, the quantum annealing algorithm proceeds forward by removing the transverse field to settle on a new classical output solution (see Figure \ref{fig:reverse_annealing}). The strength of external transverse field will determine the search to be either quasi-local or non-local. In practice, while the search is not theoretically restricted to a subset of the state space, for sufficiently small values of transverse field below the critical value, the search space will tend to be localized around the starting state. For transverse field sufficiently stronger than the critical value, the quantum fluctuations will eventually destroy the memory of the initial classical state. 

Given the ability to use reverse quantum annealing to improve known solutions, many hybrid quantum-classical algorithms become possible \cite{PatentGoogle,chancellor2017modernizing}.  One algorithm that can be hybridized with reverse annealing very naturally is a genetic algorithm; to hybridize a genetic algorithm with quantum annealing in the most straightforward way we simply use reverse quantum annealing as the mutation operator in an otherwise standard genetic algorithm.  We refer to such a genetic algorithm as a quantum-assisted genetic algorithm (QAGA).\nocite{chancellor2017modernizing,Chancellor2016b}

While many variants exist, at its essence a genetic algorithm maintains a population of solutions (individuals) that typically start as random states and evolve towards higher-quality solutions using a process that mimics evolution.  There are three main components to this process:
\begin{enumerate}
	\item Recombination (also called crossover or mating), whereby two or more individuals combine to create offspring individuals with components from each of the parent individuals.
	\item Mutation, whereby an individual is modified in a random way to create a new individual.
	\item Selection, whereby the population size is regulated by deleting certain individuals, typically those of lower quality.\footnote{Selection can also refer to the method of selecting individuals to be recombined.}
\end{enumerate}

The remainder of the paper is structured as follows.  In Section \ref{sec:qa}, we describe the Ising minimization problem,  D-Wave processors, quantum annealing, and reverse annealing.  In Section \ref{sec:simple_example}, we provide details about the first experimental implementation of a hybrid classical-quantum optimization algorithm which combines genetic algorithm with reverse annealing.  In Section \ref{sec:probing}, we test the QAGA against quantum annealing as well as established classical Ising solvers.

\section{Quantum Annealing and the Ising Minimization Problem}\label{sec:qa}

\subsection{Quantum Annealing}

D-Wave quantum processing units (QPUs) act as heuristic solvers for the \emph{Ising minimization problem}, defined on a graph $G = (V,E)$ as follows.  Given an input, or Hamiltonian, comprising a collection of fields $h = \{ h_i: i \in V \}$ and couplings $J = \{J_{ij}: (i, j) \in E \}$, assign values from $\{ -1, +1 \}$  to  $n$ {\em spin variables}  $s =\{ s_i \}$ 
so as to minimize the {\em energy function}         
\begin{eqnarray}\label{eqn:ising}
E( s ) &=  & \sum_{i\in V}  h_i  s_i   +  \sum_{(i,j)\in E} J_{ij} s_i  s_j\,.
\end{eqnarray}  The Ising minimization problem is NP-complete \cite{Barahona1982}; Ref.~\cite{Lucas2014} is a useful guide containing reductions from many NP-complete problems to Ising minimization.  Ising minimization inputs can trivially be converted to/from maximum weighted 2-satisfiability (MAX W2SAT) inputs or quadratic unconstrained binary optimization (QUBO) inputs.

Ising minimization problems can be approached using the simulated annealing (SA) metaheuristic \cite{Kirkpatrick1983}.  In simulated annealing, a random walk in the solution space is performed according to a temperature parameter---at high temperatures the random walk can easily move to higher-energy states and therefore performs a broad exploration of the solution space, whereas at low temperatures the random walk struggles to move to higher energy states and tends to settle in local optima.  If the temperature is lowered slowly enough through the course of the anneal, the random walk is able to find a ground state (global optimum) with high probability; however, this may require the cooling time to grow exponentially with the input size.

\dwave{} QPUs use a fundamentally different source of non-determinism, originating from quantum fluctuations, for heuristically finding high quality solutions of the Ising problems. The strength and the rate of change in such quantum fluctuations is controlled by a schedule known as {\em quantum annealing} (QA) \cite{Kadowaki1998}.  QA is similar in spirit to simulated annealing (SA), but rather than lowering the temperature over the course of the anneal, a \emph{transverse field}\footnote{The transverse field controls the rate of transition between states due to quantum fluctuations.} is applied strongly at first and is turned down gradually.  Meanwhile, the classical problem Hamiltonian, which is scaled down to zero at the beginning of the anneal, is gradually scaled up.  The time-dependent Hamiltonian is given by the equation:
\begin{eqnarray}
{\cal H}(s) &=& A(s)\cdot{\cal H}_{\mathrm{initial}} + B(s)\cdot{\cal H}_{\mathrm{problem}} \\
&=& A(s)\left(\sum_i \sigma_i^x\right) + B(s) \left( \sum_{i} h_i \sigma^z_i + \sum_{ij} J_{ij} \sigma^z_i \sigma^z_j \right)\,,\label{eqn:hamiltonian}
\end{eqnarray}
where $s$ is a normalized time variable ranging from $s=0$ at the beginning of a quantum anneal to $s=1$ at the end, and $\sigma_i^x$ and $\sigma_i^z$ are Pauli matrices acting on the $i^{\mathrm{th}}$ qubit.  The functions $A$ and $B$ provide the quantum annealing schedule, analogous to the cooling schedule in simulated annealing. \\

In forward annealing, the system's initial state is the uniform superposition over all states, which is a ground state of ${\cal H}_{\mathrm{initial}}$.  If the anneal is performed slowly enough in an ideal system, the system will remain in a ground state throughout the anneal; any ground state of the final Hamiltonian ${\cal H}(1)$ is a ground state of ${\cal H}_{\mathrm{problem}}$. \\

Each D-Wave QPU implements the quantum annealing algorithm in hardware and operates on Ising minimization inputs.  The graph $G$ of an input is defined by the hardwired qubits and couplers and the input Hamiltonian $(h, J)$ is programmable.  The native connectivity topology for the \dwave{} 2000Q QPU is based on a {\em Chimera graph}\footnote{A Chimera graph consists of unit tiles arranged in a square lattice, with each unit tile containing four horizontal qubits and four vertical qubits connected as a $K_{4,4}$ complete bipartite graph.  See Ref.~\cite{Chimera} for details.} containing up to $2048$ vertices (qubits) and $6016$ edges (couplers).  In each QPU, some of the qubits and couplers may be unusable due to issues in fabrication, cooldown, or calibration; the QPU used for this study had 2013 working qubits and 5818 working couplers.  Its operating temperature was approximately 12 mK (millikelvin).

\subsection{Reverse Annealing}

\begin{figure}
\centering
\graphicinput[width=3.2in]{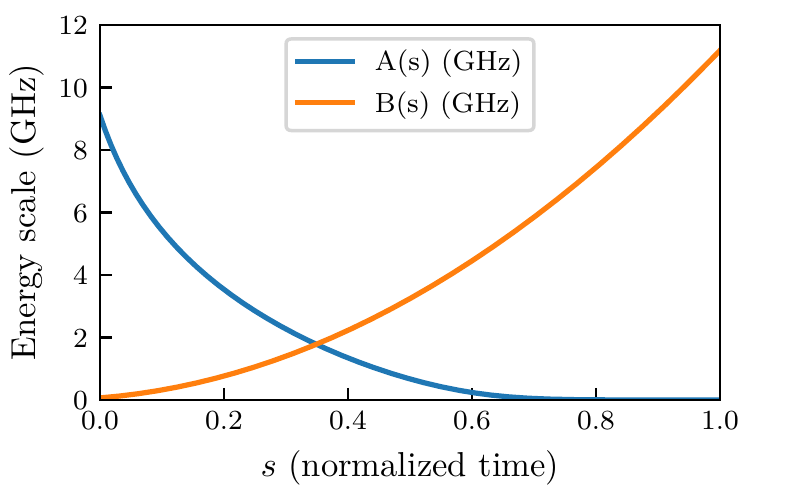}
\graphicinput[width=3.2in]{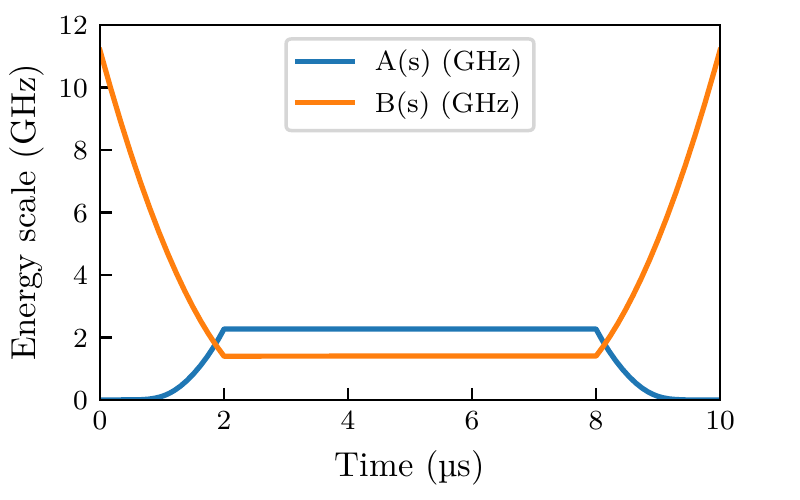}
\caption{Annealing schedules for canonical forward annealing (left) and reverse annealing (right).  In this case the reverse annealing schedule is given by the (time, $s$) pairs $((0, 1.0), (2, 0.3), (8, 0.3), (10, 1.0))$.  More explicitly, the anneal starts with the fully classical Hamiltonian ${\cal H}(1)$, evolves to ${\cal H}(0.3)$ over the first 2 \us{} of the anneal, holds the Hamiltonian ${\cal H}(0.3)$ for 6 \us{}, then spends the final 2 \us{} of the anneal evolving back to the final, fully classical Hamiltonian ${\cal H}(1)$. \label{fig:schedules}}
\end{figure}

Reverse annealing is very similar to forward annealing, though instead of starting with a quantum state (such as the uniform superposition), it starts with a classical state provided by the user.  Reverse annealing also uses a very different annealing schedule.  Rather than starting with a high transverse field $A(0)$ and a low scaling factor for the problem Hamiltonian $B(0)$, these values start at $A(1)$ and $B(1)$, respectively.  Examples of forward and reverse annealing schedules are given in Figure \ref{fig:schedules}.

In the reverse annealing schedule shown on the right of Figure \ref{fig:schedules}, the anneal starts with the fully classical Hamiltonian ${\cal H}(1)$, evolves to ${\cal H}(0.3)$ over the first 2 \us{} of the anneal, holds the Hamiltonian ${\cal H}(0.3)$ for 6 \us{}, then spends the final 2 \us{} of the anneal evolving back to the final, fully classical Hamiltonian ${\cal H}(1)$.  

We can also define this schedule with a more convenient parameterization: a 10 \us{} reverse anneal with a pause location $s^* = 0.3$ and a pause fraction of 0.6.  This means that the anneal spends 60\% of its time at the pause location with the Hamiltonian held constant as ${\cal H}(0.3)$, with the rest of its time split evenly between an evolution from ${\cal H}(1)$ to ${\cal H}(s^*)$ and an evolution back from ${\cal H}(s^*)$ to ${\cal H}(1)$.  The parameterization of (annealing time, $s^*$, pause fraction) allows us to more conveniently search the parameter space while optimizing reverse annealing and algorithms that use it.

\section{Specifying a Genetic Algorithm}\label{sec:simple_example}

\subsection{Algorithm Components}
To demonstrate a proof-of-concept for a quantum-assisted genetic algorithm, we need to specify the three components of the genetic algorithm, as well as choose appropriate inputs for demonstration.

\begin{figure}
	\includegraphics[width=\textwidth]{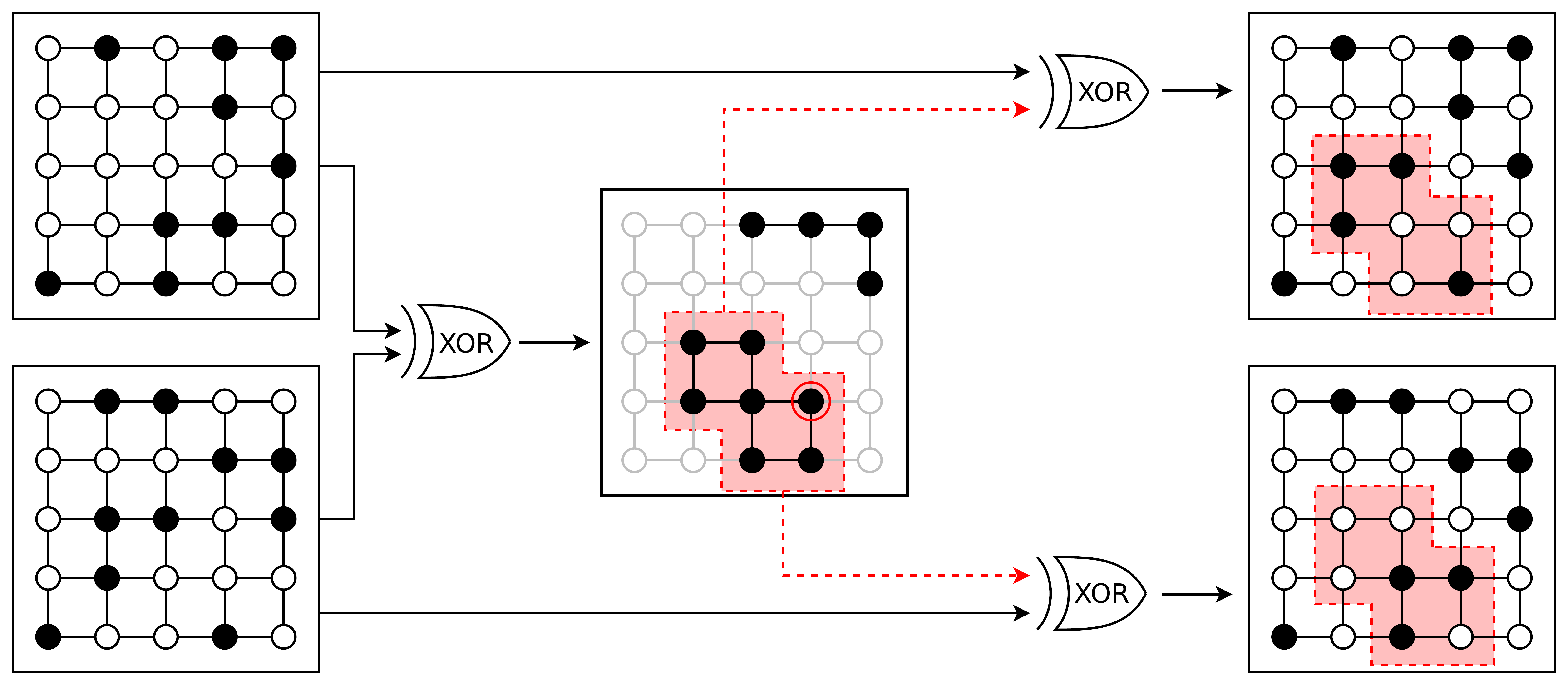}
	\caption{Example of recombination on a 2D lattice via isoenergetic cluster move.  The two ``parent'' individuals (left) are chosen and their symmetric difference ({\scshape xor} for binary states) is considered (middle).  From the symmetric difference, a variable in which the two inputs differ is selected uniformly at random (circled) and its connected component (dashed and shaded region) specifies the cluster of variables to flip.  To obtain the child individuals (right), the variables in this cluster are flipped in each of the parent individuals.\label{fig:icm}}\vspace{-3mm}
\end{figure}

\textbf{Recombination:} For recombination, we use isoenergetic cluster moves \cite{houdayer2001cluster}, also known as Houdayer moves (see Figure \ref{fig:icm}).  This recombination operation takes two individuals and randomly selects a connected cluster of spins in which the individuals differ, then flips this cluster in each individual to create two new individuals.  Isoenergetic cluster moves identify clusters of spins that would reasonably be grouped together, e.g., as a gene.  Isoenergetic cluster moves are very effective in planar or quasi-planar Ising models \cite{houdayer2001cluster,Zhu2015}.

\textbf{Mutation:} As a mutation operator, we use reverse annealing.  To mutate a state $S$, we perform a reverse anneal initialized at state $S$.  This reverse anneal applies a transverse field to evolve the classical starting state into a quantum superposition of states, then removes the transverse field to settle on a new classical state.

\textbf{Selection:} For selection, we use truncation selection; i.e., we simply keep the best $N$ individuals at the end of each generation.  This has the advantage of simplicity, but it can lead to loss of population diversity; we address this issue in Section \ref{sec:diversity}.

We implement the quantum-assisted genetic algorithm as described by the following pseudocode:

\begin{algorithm}[H]
	\begin{algorithmic}[1]
		\STATE \texttt{population} := $N$ random states
		\FOR{generation = 1 to num\_generations}
		\STATE mutate each individual in \texttt{population} with probability \texttt{mutation\_rate} 
		\STATE add the mutated states to \texttt{population}
		\STATE randomly match the individuals to make $\texttt{recombination\_rate} \times |\texttt{population}|$ pairs
		\STATE recombine each pair using an isoenergetic cluster move to make 2 new offspring
		\STATE add the offspring to \texttt{population}
		\STATE break if the algorithm has reached a stopping criterion, e.g., by timing out or reaching a certain energy
		\STATE discard individuals from \texttt{population} to maintain the desired population size of $N$
		\ENDFOR
	\end{algorithmic}
	\caption{Quantum-Assisted Genetic Algorithm}
	\label{alg:hybrid}
\end{algorithm}

\subsection{Preserving Diversity}\label{sec:diversity}

\textbf{Diversity-preserving selection.}  In early experiments we found that recombinations made up only a very small fraction of the runtime of our hybrid GA.  In principle, increasing the recombination rate should be able to improve performance with a modest runtime cost.  However, in practice, increasing the recombination rate while using simple truncation selection causes rapid loss of diversity in the population, leading to premature convergence.  This is a well-known problem in genetic algorithms and there are numerous techniques for preserving diversity.

One method of preserving diversity that fits naturally with Ising model inputs is implicit fitness sharing \cite{Smith1993}.  We use a combination of the raw Ising energy function and a shared Ising energy function; this is detailed in Appendix \ref{app:fitness}.  Our selection method preserves diversity while also ensuring that the lowest-energy solution is kept.  With this selection method in hand, we were able to improve performance by increasing the recombination rate---instead of recombining each individual with one other individual per generation, we recombined each individual with 10 individuals.  Higher recombination rates saw increased recombination costs but diminishing returns.

\textbf{Premature convergence.}  Even with diversity-preserving selection, QAGA is susceptible to premature convergence.  In other words, QAGA can get `stuck', and running for more generations is unlikely to improve solutions.  This is a well-studied issue in the area of genetic algorithms and the problem can be solved using alternative selection schemes \cite{srinivas1994adaptive,Pandey2014}.  In the interest of keeping our core algorithm simple, we instead opt to stop a run of QAGA after a fixed number of generations (in this case 50) and simply restart with a newly generated random population.

\section{Performance Comparison on Spin Glass Inputs}\label{sec:probing}

\subsection{Solvers}

We tested our quantum-assisted genetic algorithm against a suite of classical algorithms consisting of simulated annealing (SA) \cite{Kirkpatrick1983}, parallel tempering (PT) \cite{Swendsen1986}, and parallel tempering with isoenergetic cluster moves\footnote{Our implementation of PT-ICM performs isoenergetic cluster moves at every temperature; while there are more sophisticated implementations that skip the isoenergetic cluster moves at high temperatures \cite{Zhu2015}, these techniques provide only a modest constant-factor speedup while demanding more onerous tuning of parameters.} (PT-ICM) \cite{houdayer2001cluster}.
For the quantum/hybrid algorithms, we use quantum annealing (QA) and our quantum-assisted genetic algorithm (QAGA). We optimized parameters for each of the five solvers; details are given in Appendix \ref{app:parameters}.

\subsection{Bookkeeping}\label{sec:bookkeeping}

This manuscript aims to demonstrate that a quantum-assisted genetic algorithm is a viable option in the realm of hybrid quantum/classical algorithms, and that quantum-assisted genetic algorithms bear further research and development. To facilitate our development and analysis, we use simple runtime models to estimate the amount of time required for each classical operation.    
This makes it possible to compare quantum and classical components in a larger hybrid algorithm.  The alternative to using simple runtime models is to measure wall-clock times; however, for our prototype implementations that call the QPU synchronously, wall-clock time is dominated by network latency, waveform generation, and QPU initialization time, which are not the focus of this study. To simplify our development and analysis we account for computational effort as follows:

\begin{itemize}
	\item For Metropolis or Gibbs sweeps involved in SA, PT, and PT-ICM, we define the cost as 0.2\,ns per spin flip update; for the $\sim$2000-qubit inputs considered in this section, this is roughly 0.4\,\us{} per sweep.
	\item For isoenergetic cluster moves either used in PT-ICM, or used as a recombination operator in QAGA, we define the cost as 0.2\,ns per spin flip update; for the $\sim$2000-qubit inputs considered in this section, this is roughly 0.4\,\us{} per isoenergetic cluster move.  The work involved in this step is linear in the number of variables since we must identify the variables in which two individuals differ.  Our cost model is consistent with experimental results.
	\item For reverse anneals, we define the cost as the total annealing time (including pause time),
discounting QPU programming time, readout time, and other overhead.
	\item For PT and PT-ICM, we spend considerable computational effort to optimize the temperatures used; we do not include this parameter optimization effort in our costs.
\end{itemize}

For the classical algorithms, the runtime is dominated by the Metropolis/Gibbs sweeps.  Our cost model is consistent with timings reported for highly optimized solvers \cite{Isakov2014a}.

\subsection{Input Classes}

In this study we analyzed three different input classes, each generated randomly based on the underlying graph of the D-Wave QPU and its functional qubits used for these experiments.  All inputs were generated based on the full chip, meaning that they all have roughly 2000 variables.  All three of these input classes have fields (i.e., $h$ values) set to zero.

\textbf{RAN1---Bimodal spin glasses.}  These inputs are generated by randomly assigning each coupling a value from $\{-1, 1\}$.  These inputs have been studied extensively as inputs to D-Wave QPUs \cite{Roennow2014,Selby2014,King2016floppy,King2015,Steiger2015} because they are nontrivial optimization problems that are trivial to generate.  The low precision of RAN1 inputs, along with the fact that they are generated based on the physical layout of the QPU, work in the favor of D-Wave QPUs.  However, high degeneracy of first-excited states can make it extremely hard for quantum annealers to find ground states for these inputs \cite{King2016floppy}.

\textbf{AC3---Anticluster inputs.}  The inputs we studied for this section were randomly generated AC3 problems.  AC3 problems are random spin-glass inputs whose inter-tile couplings are multiplied by a factor of 3.  This works to combat the quasi-planarity of the Chimera topology \cite{Katzgraber2014}.  D-Wave QPUs perform well on these inputs relative to classical solvers when it comes to finding good approximate solutions \cite{King2015}; however, at the 2000-qubit scale, D-Wave QPUs struggle to find global optima, as we demonstrate in this section.

\textbf{DCL---Deceptive cluster loops.}  Deceptive cluster loop inputs \cite{Mandra2018} are a variant of frustrated cluster loop (FCL) inputs \cite{King2019}.  FCL and DCL inputs are generated by creating dense clusters of qubits that are treated as logical variables in a constraint satisfaction problem generated from frustrated loops \cite{Hen2015,King2015range}.  These inputs are designed to depend on quantum tunneling locally, and to require the solution of a constraint satisfaction problem globally.  When variables in a cluster are coupled weakly enough that the ground state of the input may contain broken clusters, the clusters are considered deceptive because they can fool logical solvers.  We generated the inputs per Ref.~\cite{Mandra2018} with parameters $(\alpha, R, \lambda) = (0.75, 4.0, 4.0)$.

\subsection{Results}

\newlength{\imheight}
\setlength{\imheight}{6cm}

\begin{figure}
	\begin{tabular}{ccc}
		\includegraphics[height=\imheight]{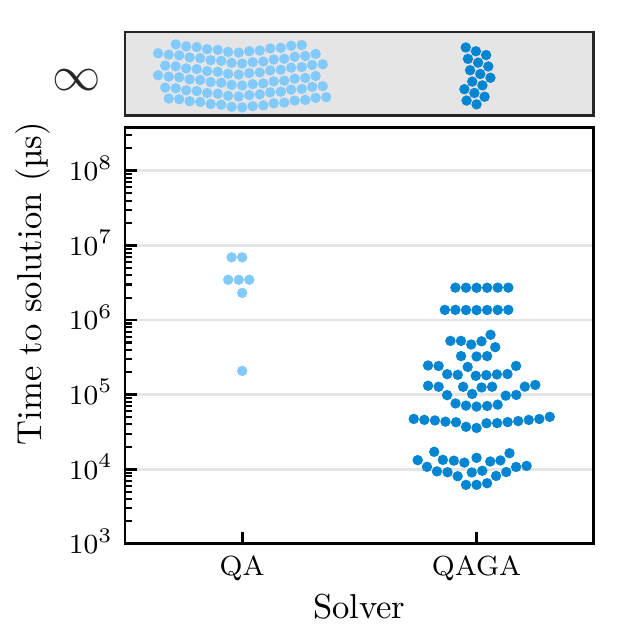} & \includegraphics[height=\imheight]{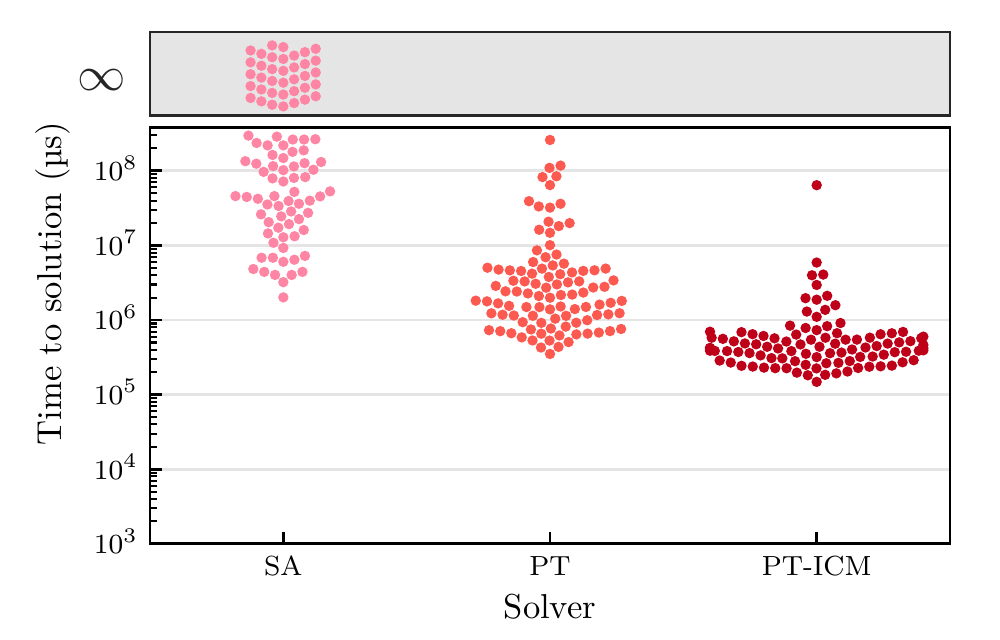} & \hspace{-1mm}
		\rotatebox{90}{\Large ~~~~~~~~~~~~~~~~RAN1} \vspace{-2mm} \\ 
		\includegraphics[height=\imheight]{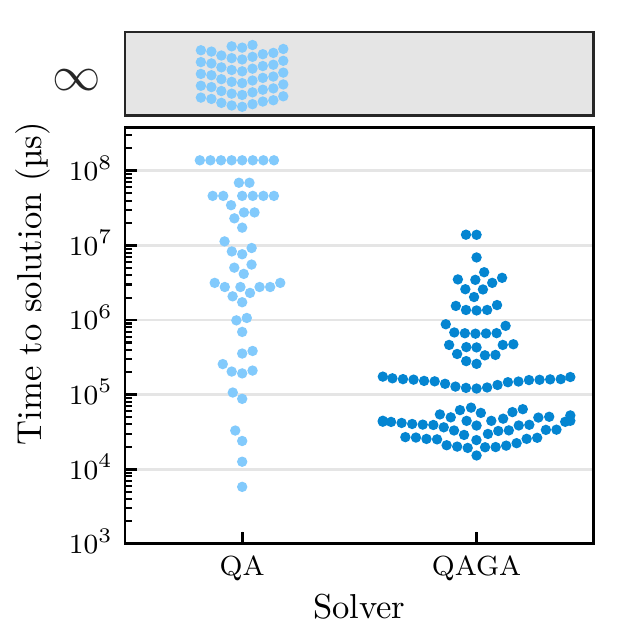} & \includegraphics[height=\imheight]{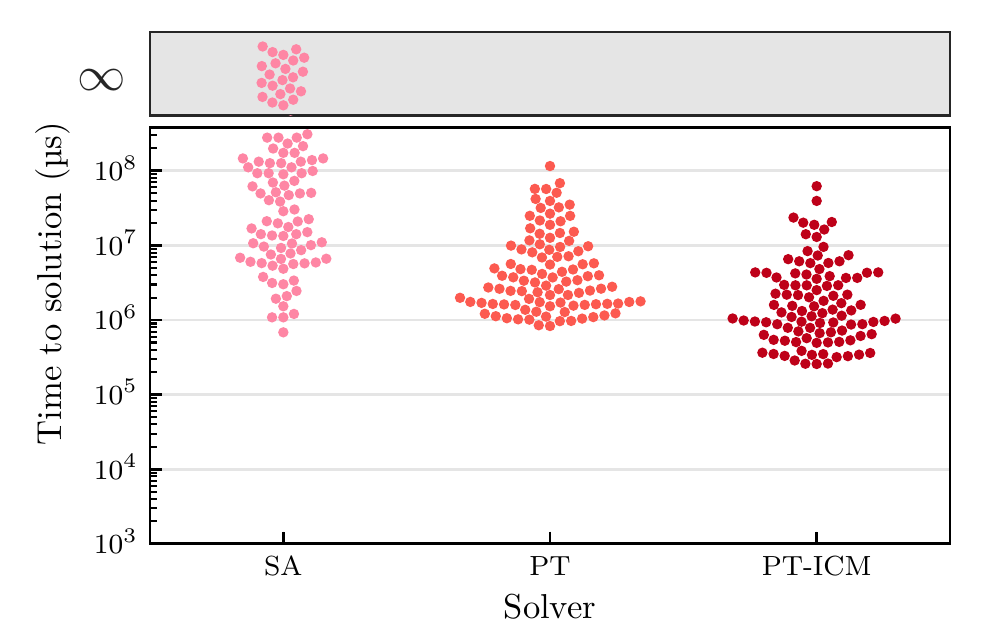} & \hspace{-1mm}
\rotatebox{90}{\Large ~~~~~~~~~~~~~~~~AC3} \vspace{-2mm} \\ 
		\includegraphics[height=\imheight]{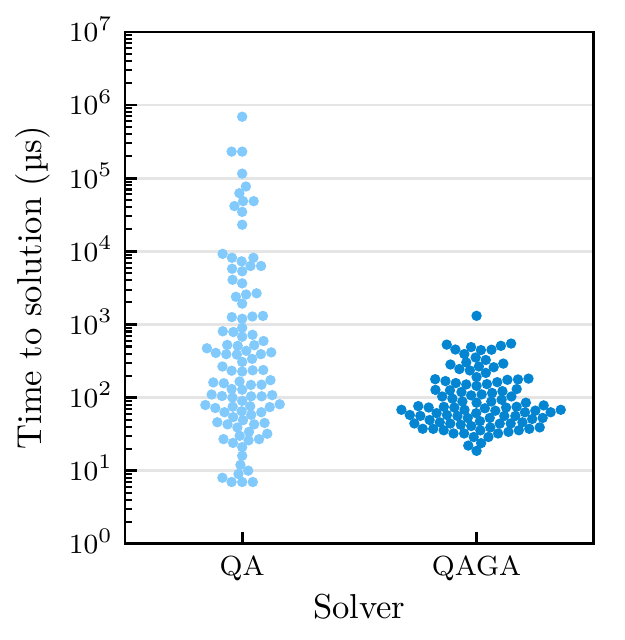} & \includegraphics[height=\imheight]{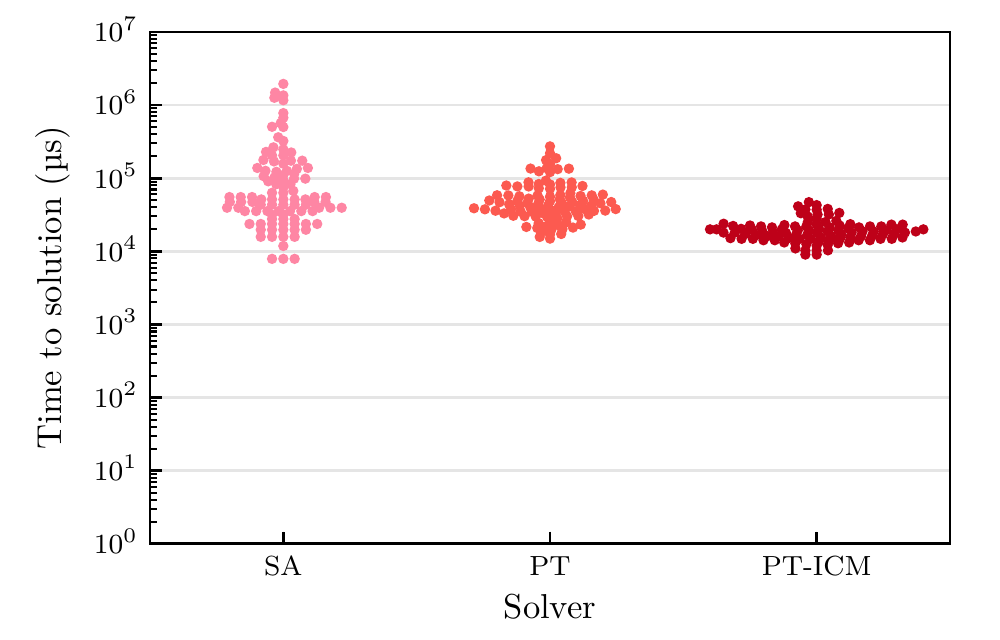} & \hspace{-1mm}
\rotatebox{90}{\Large ~~~~~~~~~~~~~~~~DCL} \vspace{-2mm} \\ 
	\end{tabular}\caption{Performance of QA, QAGA, SA, PT, and PT-ICM on the three input classes studied.  Each plot is a swarm plot in which a dot represents a single input/solver pair.  The $y$-axis shows the median time-to-solution (TTS) for the given solver on the given input, where the median is calculated over repeated runs.  Results are based on a simplified model of minimal runtime (see Section \ref{sec:bookkeeping}).  Plots are split into quantum and hybrid algorithms (left) and classical algorithms (right) because the difference between our reported time and wall clock time is much greater for the quantum and hybrid algorithms.  In all three classes, QAGA had a much tighter distribution of TTS, mitigating the tendency of QA to perform very poorly on a small fraction of the inputs.  Note that RAN1 was the hardest problem class for the solvers, followed by AC3, then DCL inputs which were significantly easier for all solvers.  Comparing QAGA to the classical solvers, QAGA's performance was dominant on the AC3 and DCL inputs.  On RAN1 inputs, QAGA performed better on most inputs, but failed to reach ground states on 16 of the 100 inputs, whereas PT and PT-ICM were able to find ground states for all 100.\label{fig:all_swarm}}
\end{figure}

For each problem class, we generated 100 random inputs and ran the five solvers (QA, QAGA, PT, PT-ICM, SA) on all inputs.  The putative ground state energies were determined and corroborated using many independent long runs of PT-ICM.

While we attempted to run all solvers for long enough to find ground states (and therefore estimate the time to solution (TTS)), this was impractical for some of the solvers and some of the inputs.  We gave each solver roughly five minutes of computation time on each input (to be precise, $10^{8.5}$ \us{}).  Some RAN1 inputs caused SA, QA, and QAGA to time out, and some AC3 inputs caused SA and QA to time out.

Figure \ref{fig:all_swarm} shows results of QA and QAGA, providing a comparison of the TTS distributions of the two solvers on the 100 inputs from each class.  These plots show that QAGA has a significantly tighter distribution of time-to-solution.  Most importantly, the TTS distribution for QAGA appears to have a much smaller (or skinnier) upper tail than the distribution for QA (see Refs. \cite{Steiger2015} and \cite{King2016floppy} for detailed discussion of heavy tails in TTS distributions for QA on RAN1 problems).

\setlength{\imheight}{7cm}

\begin{figure}
	\begin{tabular}{ccc}
		{\Large ~~~~~QAGA vs.~QA} & {\Large ~~~~~QAGA vs.~PT-ICM} & \\
		\includegraphics[height=\imheight]{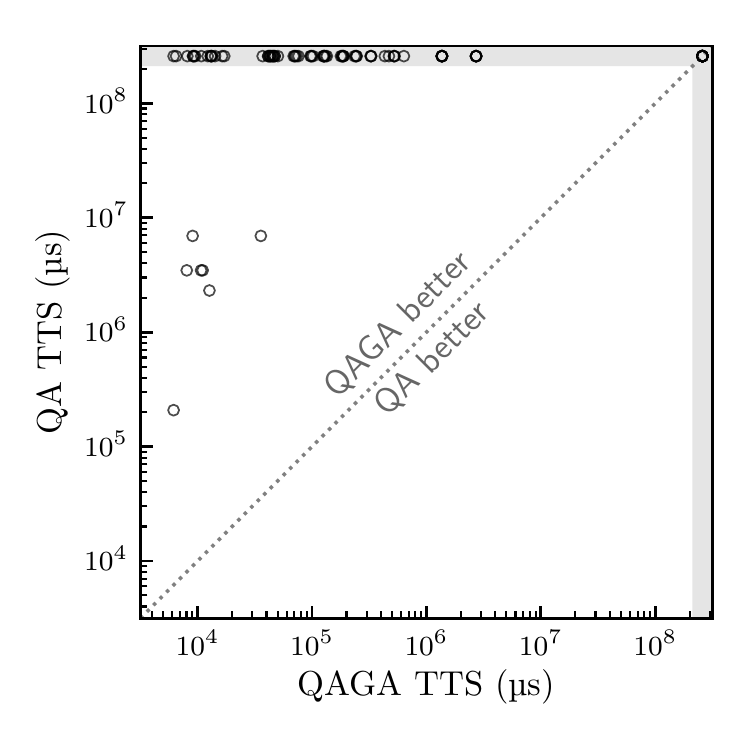} &
		\includegraphics[height=\imheight]{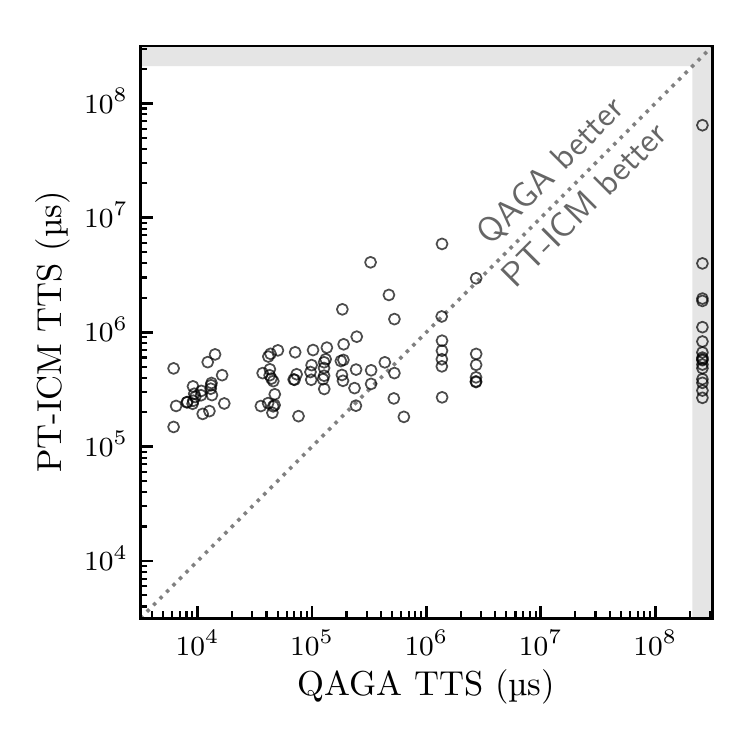} & 
		\rotatebox{90}{\Large ~~~~~~~~~~~~~~~~~~~~~RAN1} \vspace{-4mm} \\ 
		\includegraphics[height=\imheight]{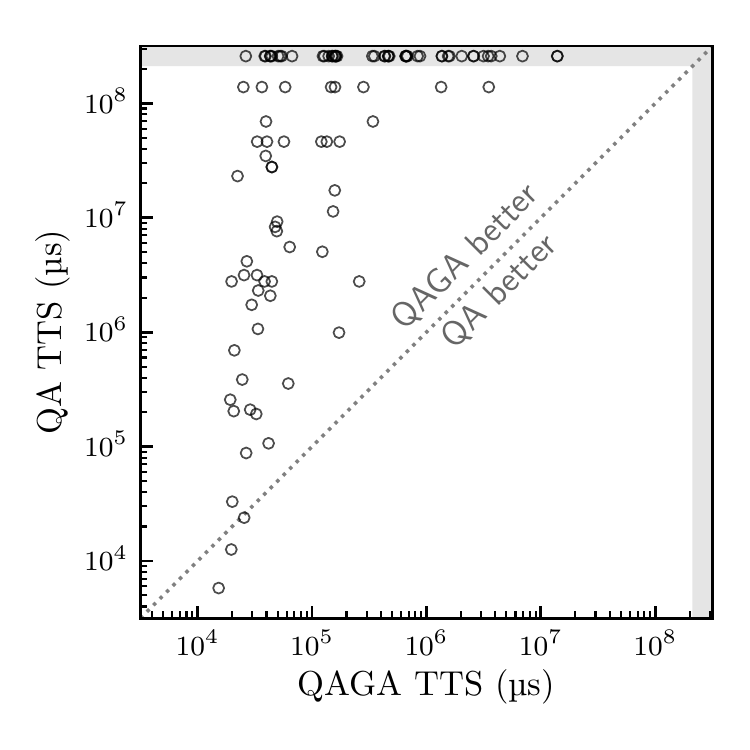} &
		\includegraphics[height=\imheight]{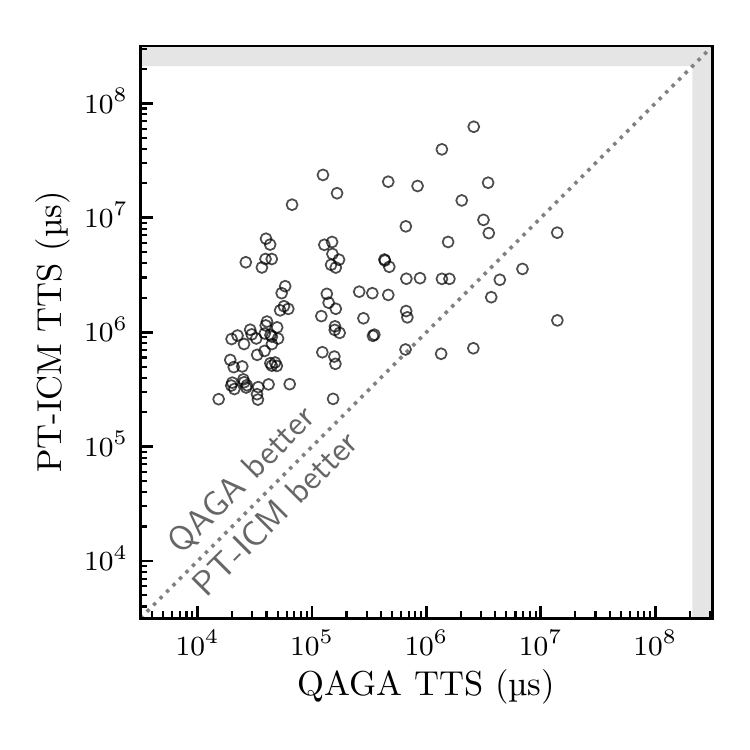} &
		\rotatebox{90}{\Large ~~~~~~~~~~~~~~~~~~~~~AC3} \vspace{-4mm} \\ 
		\includegraphics[height=\imheight]{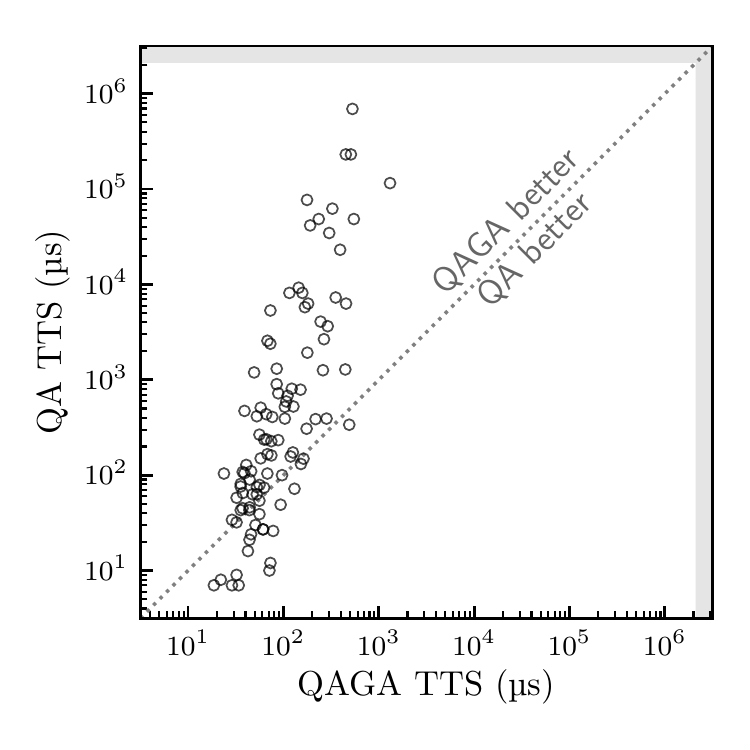} &
		\includegraphics[height=\imheight]{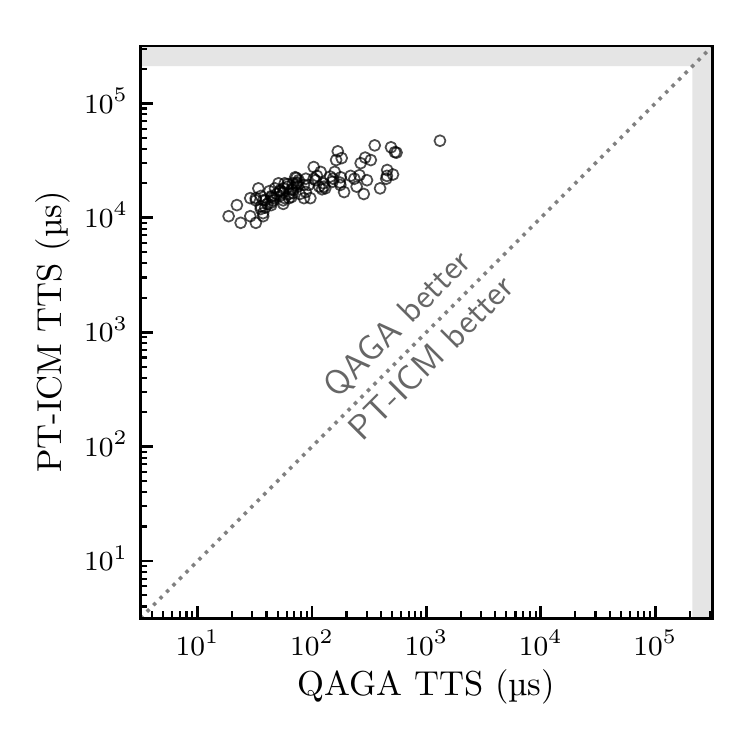} &
		\rotatebox{90}{\Large ~~~~~~~~~~~~~~~~~~~~~DCL} \\
	\end{tabular}\caption{Head-to-head comparisons of QAGA versus other solvers.  Comparisons against QA are in the left column and comparisons against PT-ICM are in the right column.  In each scatter plot, one circle represents a single input.  The shaded region in each plot indicates TTS that is estimated as infinite because a ground state was never reached.  Results are based on a simplified model of minimal runtime (see Section \ref{sec:bookkeeping}).\label{fig:head_to_head}}
\end{figure}

Figure \ref{fig:head_to_head} provides scatter plots that show head-to-head comparisons of QAGA versus the two solvers of greatest interest: QA (which serves as a baseline quantum algorithm) and PT-ICM (which is the best known classical algorithm for these inputs).  The left column of these plots, which shows performance of QAGA versus QA, confirms a point that is suggested in Figure \ref{fig:all_swarm}.  Namely, that QA only outperforms QAGA on the easiest of the random inputs.

In the right column of Figure \ref{fig:head_to_head}, we see the comparison between QAGA and PT-ICM.  We can see that the relative performance of these two solvers is highly dependent on the input class:
\begin{itemize}
\item DCL inputs were easy for QAGA but still somewhat challenging for PT-ICM, with QAGA beating PT-ICM by over two orders of magnitude on average.
\item AC3 inputs were more challenging for both solvers.  QAGA outperformed PT-ICM by roughly an order of magnitude on average, with PT-ICM being faster on only 7 out of 100 inputs.
\item RAN1 inputs were the hardest for QAGA but easier for PT-ICM than AC3 inputs.  While QAGA performed better on most of the inputs, the time-to-solution distribution for QAGA had a longer tail, with QAGA timing out on 16 out of 100 inputs.  30 of the 100 inputs took over 1\,s for QAGA, whereas only 12 of the 100 inputs took over 1\,s for PT-ICM.
\end{itemize} 

\textbf{Wall-clock time.} The main caveat regarding these results is that, while all of our solvers' actual runtimes are slower than the hypothetical ideal runtimes reported, the difference is most extreme for our quantum-assisted genetic algorithm.  There are two main types of overhead associated with the use of the QPU in a hybrid algorithm: overhead that requires exclusive access to the QPU (e.g., programming and readout time) and overhead incurred ``outside'' the QPU (e.g., network latency and conversion of an input Hamiltonian to analog waveforms).  Both types of overhead can be reduced in the coming years through engineering improvements and by bringing the CPU or GPU running the outer genetic algorithm closer to the QPU.

Because QPU overhead will never be eliminated completely, and because the quantum annealing time itself is non-negligible, near-term hybrid algorithms will need to be asynchronous to optimize performance as measured by wall-clock times.  After a mutation call is sent to the QPU, the classical part of the algorithm will continue working while it waits for the QPU response to return.  How exactly this should be performed is an area of active research.

\section{Conclusions}

We have implemented and demonstrated the utility of a quantum-assisted genetic algorithm that uses reverse quantum annealing as a mutation operator.  The results are promising, suggesting that reverse annealing is a viable mutation operator and that variants of QAGAs, or more generally, hybrid quantum-classical population-based algorithms, are a potential avenue to computational advantage. It would be conceivable that future highly-competitive heuristic optimization platforms would be required to employ quantum and classical co-processor \cite{PatentGoogle} to harness the complementary nature and the interplay of quantum and classical fluctuations for faster sampling of the solution space.

\bibliography{paper}

\appendix

\section{Shared Ising Energy Function}\label{app:fitness}

\textbf{Shared Ising energy.}
When using a high recombination rate, we use a form of implicit fitness sharing \cite{Smith1993} to maintain population diversity.  In implicit fitness sharing, each ``reward'' component of the fitness function is shared evenly among all of the individuals that receive that reward.  Using a shared Ising energy function promotes diversity by giving more weight to good characteristics that are rare in the population.  For Ising models, a reward will be either a field $h_i$ that is satisfied or a coupling $J_{ij}$ that is satisfied.  In this study, all fields $h_i$ are set to zero in all inputs, so all rewards are satisfied couplings.

For example, if we have a population of 40 states, and there is a coupling $J_{ij}=1$ that is satisfied by 5 of our 40 states, that coupling will contribute an energy of $-0.2$ to the shared Ising energy of each of those 5 states, and will not affect the shared Ising energies of the other 35 states in our population.  If there is another coupling $J_{k\ell}=-0.5$ that is only satisfied by one of our 40 states, it will contribute $-0.5$ to that state's shared Ising energy and nothing for the other 39 states.

To define a shared Ising energy function more explicitly, let $n_i$ be the number of individuals for which $h_i s_i$ is negative, and let $n_{ij}$ be the number of individuals for which $J_{ij} s_i  s_j$ is negative.  Further, we define 

$$
f_i = \begin{cases} \frac{h_i  s_i}{n_i} &\mbox{if } h_i s_i < 0  \\ 
0 & \mbox{otherwise. } \end{cases}
$$

$$
f_{ij} = \begin{cases} \frac{J_{ij} s_i  s_j}{n_{ij}} &\mbox{if } J_{ij} s_i  s_j < 0  \\ 
0 & \mbox{otherwise. } \end{cases}
$$

With these definitions in hand, the formula for shared Ising energy is simply:

$$
E_{\mathrm{shared}}( s ) = \sum_{i\in V}  f_i   +  \sum_{(i,j)\in E} f_{ij}\,.
$$

By contrast, in (explicit) fitness sharing, individuals that are closer than some \emph{sharing radius} $R$ have their fitnesses reduced to represent competition for resources.  We use implicit fitness sharing because it is simpler for our use case---implicit fitness sharing does not require the tuning of a sharing radius.

\textbf{Pareto selection.}
Using a shared Ising energy function alone for selection would be problematic, since the objective of our optimization is Ising energy rather than shared Ising energy.  In order to ensure that the best states are kept in our population\footnote{This property is known as \emph{elitism}.} while also maintaining diversity, we use a selection method involving two fitness functions.  Specifically, we consider the Pareto frontier, a concept that is central to the field of multi-objective optimization (see, e.g., \cite{konak2006multi}).

The Pareto frontier is the set of solutions such that, if a solution $x$ is in the set, no other solution is better than $x$ in \textit{both} metrics.  We order the elements on the Pareto frontier, considering the primary metric to be more important than the secondary metric.  When the Pareto frontier is exhausted, we remove those solutions from the set being ordered and consider the Pareto frontier of the remaining solutions, effectively peeling a layer off of the two-dimensional point set.  We continue iterating through the solutions in this manner until all have been ordered.

The resulting order is used to determine which solutions are kept and which are discarded---if our algorithm requires us to keep $k$ solutions, we simply keep the first $k$ according to the order described above.

The two fitness functions used in our selection method are raw Ising energy (i.e., the energy according to the input Hamiltonian), and shared Ising energy (i.e., the energy calculated using implicit fitness sharing).  Each solution in the population is graded using raw Ising energy as the primary metric and shared Ising energy as the secondary metric.  These two metrics are ways of quantifying, respectively, quality and valuable uniqueness.

\section{Hyper-parameter Optimization}\label{app:parameters}

In this section we discuss the hyperparameters that were optimized for each solver.  To provide concrete examples, we give parameter values chosen for the AC3 problem class, though the parameters were optimized separately for each problem class.  Parameters for each class were optimized on a set of 30 inputs which were not used in our main experiments.

\paragraph{Quantum annealing}

The only parameter we optimized for QA was annealing time.  We tried values of 10 \us{}, 100 \us{}, and 1000 \us{}.  Overall TTS performance was roughly the same for 10 \us{} and 100 \us{}, but significantly worse for 1000 \us{} anneals.  We opted to run the tests with 100 \us{} anneals because, relative to 10 \us{} anneals, using 100 \us{} anneals allows us to use a greater fraction of wall-clock time on actual annealing (i.e., on computation rather than overhead), making it possible to find ground states on harder inputs.

\paragraph{Quantum-assisted genetic algorithm}

The quantum-assisted genetic algorithm has annealing parameters as well as standard genetic algorithm parameters.  The annealing schedule used was crafted through trial and error on a small number of inputs; we settled on the schedule given by the (time, $s$) pairs ((0.0, 1.0), (1.0, 0.5), (7.0, 0.5), (10.0, 1.0)).  In other words, we use a reverse anneal of 10 \us{} total annealing time, including a 1 \us{} evolution from $s=1.0$ to $s=0.5$, a 6 \us{} dwell at $s=0.5$, then a 3 \us{} evolution from $s=0.5$ to $s=1.0$.

A population size of 40 was selected, along with a mutation rate of 1 mutation per individual.  These were chosen primarily to align with the restrictions of a prototype feature known as batch reverse annealing.  This feature allows reverse annealing from multiple initial states in a single API call; in this case the limit was 40 initial states in a single call.  This feature accelerated experimentation significantly and it was therefore of practical interest to work within its limitations, i.e., to have a maximum of 40 mutations per generation.  We tried using smaller populations but performance suffered.  It is possible to use larger populations with a lower mutation rate, but we will save this parameter exploration for future work.

In addition to the fitness sharing method of preserving genetic diversity in the population, at the end of each generation we kept only 30 states, with 10 new random states being added to the population at the end of a generation.  This was found to improve performance significantly.

The recombination rate was optimized and a value of 10 was chosen so that each individual would be in 10 recombinations per generation.

\paragraph{Parallel tempering}

In parallel tempering, the parameters required to define the algorithm are simply the beta values, i.e., inverse temperatures, that define the replicas.  For each input we pre-selected betas that yielded exchange rates roughly between 0.3 and 0.8.  The number of temperatures used varied slightly across the inputs, but averaged around 60.

Note that preselecting parameters for each input goes against proper experimental methodology and could unfairly benefit the three classical solvers that use these temperatures.  However, if we had selected parameters for a distribution of inputs, the parameters would likely lead to suboptimal performance on certain inputs.  To guarantee that we are not disadvantaging classical solvers, we chose to preselect parameters for the classical solvers on a per-input basis.

\paragraph{Simulated annealing}

For simulated annealing, we need to provide a temperature schedule, including the total number of sweeps per anneal.  In this case we performed a grid search for the optimum number of sweeps per anneal, given that the schedule of betas is a linear interpolation of the optimal betas found for parallel tempering.  The optimal number of sweeps per anneal was found to be $10^5$.

\paragraph{Parallel tempering with isoenergetic cluster moves}

For PT-ICM we can use the same temperatures we used for parallel tempering.  The one additional parameter to optimize is the frequency of cluster moves.  We found that performing cluster moves every 3 sweeps was approximately optimal.

\section{Time-to-Solution Calculation}

The performance metric used in this study is median TTS, i.e., $\mathrm{TTS}_{50}$.  We chose to analyze the median instead of distributional tail percentiles such as $\mathrm{TTS}_{95}$ or $\mathrm{TTS}_{99}$ that are frequently used because the median is a far more stable statistic even for distribution-free estimates, i.e., sample medians.  In certain cases, such as with simulated annealing where each sample can be considered a fixed-time Bernoulli trial, it is possible to define a simple model for the runtime distribution that yields reliable estimates of tail percentiles.  However, we are not aware of any such simple parametric model for PT and PT-ICM.

\paragraph{Simple annealers}

For simple annealers, i.e., simulated annealing and quantum annealing, we model time to solution using a geometric distribution.  With a probability $p_s$ of returning a global optimum in any given sample, and for a runtime $t_a$ of a single anneal, this gives $$\mathrm{TTS}_{50} = t_a \left\lceil\frac{-1}{\log_2(1-p_s)}\right\rceil\,.$$

We performed $10^6$ anneals for simulated annealing and $2\cdot10^6$ anneals for quantum annealing.

\paragraph{Quantum-assisted genetic algorithm}

For the quantum-assisted genetic algorithm the $\mathrm{TTS}_{50}$ calculation is similar but slightly different.  The runtime per generation is fixed.  Additionally, we can model the number of random restarts required (i.e., the number of times QAGA hits the maximum of 50 generations without finding a global optimum) using a geometric distribution.

Since we limit the number of generations we run for, we first model the number of restarts required to reach a global optimum using a geometric distribution, noting that, when the maximum number of generations is reached without finding a global optimum, the number of generations, and therefore the runtime, is fixed.  We then add the sample median of the number of generations required when the generation limit is not hit, where this sample median is taken over all successful QAGA runs using given parameters on a given input.

Using the optimized parameter set, we performed 200 runs of QAGA on each input, with each run consisting of at most 50 generations.

\paragraph{Tempering methods}

For PT and PT-ICM, no simple parametric model exists for the runtime.  We therefore simply use the sample median of 100 independent runs.  On each run, either PT or PT-ICM was run with optimized parameters until a ground state was reached.

\end{document}